\documentclass[10pt,letterpaper,twocolumn]{article} 

\usepackage{ol2}
\usepackage[draft]{hyperref}
\usepackage{amsmath}

\begin{document}

\twocolumn[ 

\title{Quantum simulation of decoherence in optical waveguide lattices}


\author{Stefano Longhi}

\address{Dipartimento di Fisica, Politecnico di Milano and Istituto di Fotonica e Nanotecnologie del Consiglio Nazionale delle Ricerche, Piazza L. da Vinci 32, I-20133 Milano, Italy (stefano.longhi@polimi.it)}

\begin{abstract}
We suggest that propagation of nonclassical light in lattices of optical waveguides can provide a laboratory tool to simulate quantum decoherence phenomena with high non-Markovian features.  As examples, we study decoherence of optical Schr\"{o}dinger cats in a lattice that mimics a dissipative quantum harmonic oscillator coupled to a quantum bath, showing fractional decoherence in the strong coupling regime, and Bloch oscillations of optical Schr\"{o}dinger cats, where damped revivals of the coherence can be observed. 
\end{abstract} 



 ] 

\noindent 
Lattices of evanescently-coupled optical waveguides probed with either classical or non-classical states of light have provided over the past decade a useful laboratory tool
for simulating a wealth of coherent quantum phenomena \cite{R1,R2,R3,BO1,BO2,BO3,BO4,DL,AL1,AL2,AL3,Z1,Z2,QW1,QW2,QW3,QW3bis,QW4}, including Bloch oscillations and Zener tunneling \cite{R1,BO1,BO2,BO3,BO4}, dynamic \cite{DL} and Anderson \cite{AL1,AL2,AL3} localization, quantum Zeno effect \cite{Z1,Z2}, and quantum walks \cite{QW1,QW2,QW3,QW3bis,QW4} to mention a few. The effective decoherence-free properties of photons in waveguide lattices make them a very attractive test-bed for realizing many quantum mechanical behaviors that require low decoherence. On the other hand, decoherence 
of quantum objects coupled to the environment plays a major role in quantum science and technology, for example it is at the heart of the quantum-classical boundary \cite{decoh,Harochebook}.  Cavity quantum electrodynamics experiments \cite{QE1,QE2} beautifully showed decoherence of nontrivial photon states, including Schr\"{o}dinger cat states. Here decoherence is generally observed as an irreversible  Markovian  process, which can be described by a Lindblad-type master equation for the reduced density matrix \cite{Harochebook}. In recent works an increasing attention has been devoted to the investigation of  memory (non-Markovian) effects in decoherence processes \cite{NM1,NM2,NM3,NM4,frac0,frac} and to the realization of quantum simulators where the decoherence dynamics can be measured and controlled \cite{NM4,trapped}.\par
In this Letter we suggest that lattices of optical waveguides probed with non-classical states of light can provide an attractive test bed for simulating decoherence phenomena in highly non-Markovian regimes. This setup can offer several advantages, including the possibility to simulate ultrastrong coupling regimes leading to limited decoherence \cite{frac0,frac} and to 
perform  statistical measurements
on different realizations without the need to store photons in cavities. In our photonic realization, the open quantum system is provided by a reference waveguide W of the array, whereas the surrounding waveguides provide the structured quantum bath B into which photons can decay via evanescent coupling \cite{Z1,Z2}. Photon state propagation along the longitudinal direction $z$ of the array emulates a unitary evolution in time $t$ described by a system-bath  Hamiltonian $\hat{H}$, with $t=z/c$. We will specifically discuss the photonic simulation of two non-Markovian effects:  revivals and fractional decoherence \cite{frac0,frac} of a dissipative quantum oscillator in the strong coupling regime, and damped revivals of coherence for an optical Schr\"{o}dinger cat state \cite{cat} undergoing Bloch oscillations (BOs) and Zener tunneling (ZT).\\ 
As a first example, we  propose a waveguide simulator of the dissipative harmonic oscillator, which is a workhorse for the study of decoherence in quantum mechanics and quantum optics \cite{Harochebook,Mil}. The optical structure, shown in Fig.1(a), consists of an optical waveguide W side-coupled to a semi-inifinite waveguide array, which provides a structured reservoir B \cite{Z1,Z2}.  Assuming nearest-neighbor coupling from the evanescent fields between waveguides $j$ and $j+1$ (coupling constant  $\kappa_j$) and propagation constant $\sigma_j$,  in the tight-binding approximation the Hamiltonian  $\hat{H}$ (with $\hbar=1$) reads \cite{QW1,QW2,QW3}
\begin{equation}
\hat{H}=\sum_{j=0}^{\infty}  \sigma_j \hat{a}^{\dag}_{j} \hat{a}_{j}+ \sum_{j=0}^{\infty} \left( \kappa_{j} \hat{a}^{\dag}_{j} \hat{a}_{j+1}+ {\rm H.c.} \right)
\end{equation}
where $\hat{a}^{\dag}_j$ and $\hat{a}_j$ are the bosonic creation and annihilation
operators for a photon in waveguide $j$ and $j=0$ is the boundary defective waveguide W. We assume a uniform semi-infinite array B with hopping rate $\kappa_j=\kappa$ and $\sigma_j=0$ for $j \geq 1$, and a defective waveguide W with propagation constant detuning $\sigma_0$  coupled to the semi-array B with a hopping rate $\kappa_0$. Memory effects arise in the strong coupling  regime $\kappa \sim \kappa_0$, whereas a Markovian dynamics is attained in the weak coupling limit $\kappa_0 \ll \kappa$ \cite{Z1,Z2}. To show the equivalence of the Hamiltonian (1) with that of a dissipative quantum harmonic oscillator \cite{Harochebook,Mil}, it is worth introducing the bosonic creation operators  $\hat{b}^{\dag}(q)$ of Bloch modes in the semi-infinite array B via the relation \cite{QW3} $\hat{b}^{\dag}(q)= (2/ \pi)^{1/2} \sum_{j=1}^{\infty}\hat{a}^{\dag}_j \sin (jq)$, where $ -\pi \leq q < \pi$ is the Bloch wave number.  In this way the Hamiltonian (1) takes the form
\begin{equation}
\hat{H}= \sigma_0 \hat{a}^{\dag}_{0} \hat{a}_{0}+\int_{-\pi}^{\pi} dq \; \omega(q) \hat{b}^{\dag}(q) \hat{b}(q)+ \hat{H}_{int}
\end{equation}

\begin{figure}[htb]
\centerline{\includegraphics[width=7.6cm]{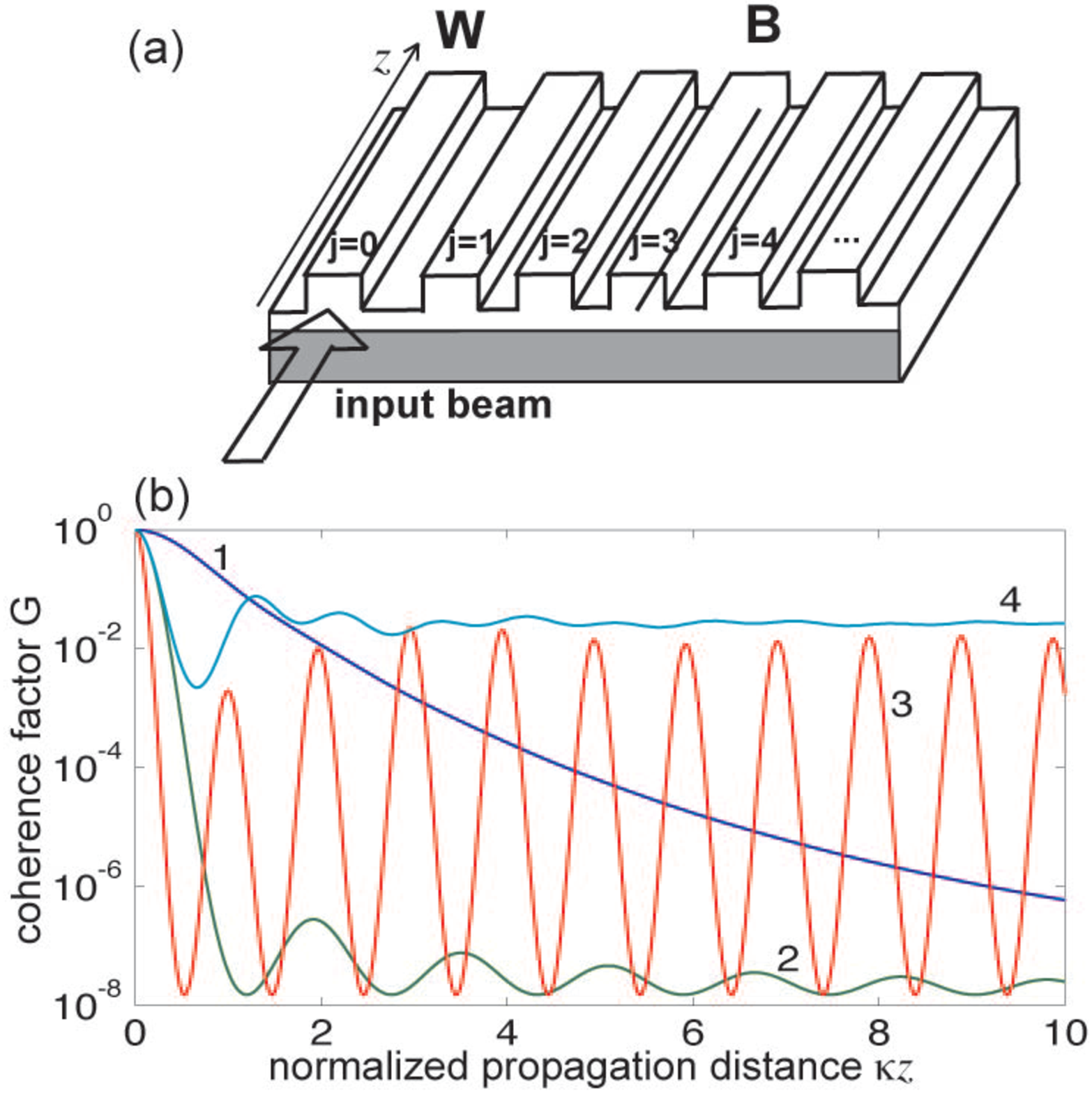}} \caption{
(Color online)  (a) Schematic of a waveguide W side-coupled to a semi-infinite homogeneous array B for the simulation of a dissipative harmonic oscillator.  The waveguide W is excited at the input plane by a Schr\"{o}dinger cat state. (b) Behavior of the coherence factor $G(z)$ (in logarithmic scale) versus normalized propagation distance $\kappa z$ for: $\kappa_0 / \kappa=0.2$, $\sigma_0/ \kappa=0$ (curve 1);  $\kappa_0/\kappa=\sqrt{2}$, $\sigma_0/ \kappa=0$ (curve 2);  $\kappa_0/ \kappa=3$, $\sigma_0/  \kappa=0$ (curve 3); and $\kappa_0/\kappa=\sqrt{2}$, $\sigma_0 / \kappa=4$ (curve 4). }
\end{figure}
were we have set $\omega(q)=2 \kappa \cos q$, $\hat{H}_{int}=\int_{-\pi}^{\pi} dq \{  g(q) \hat{a}^{\dag}_0\hat{b}(q)+ {\rm H.c.} \}$ and $g(q)=\kappa_0 (2/\pi)^{1/2} \sin q$. In this form, the Hamiltonian $\hat{H}$ describes a quantum harmonic oscillator of frequency $\sigma_0$, coupled to a bath of harmonic oscillators of frequencies $\omega(q)$ with a colored interaction $g(q)$. Let us assume that at the $z=0$  input plane the boundary waveguide W is excited by an optical Schr\"{o}dinger cate state \cite{cat}, i.e. let us assume that the state vector of the system at $z=0$ is given by 
\begin{equation}
| \mathcal{Q}(z=0) \rangle=\left( 1 / \sqrt{ \mathcal{N}} \right) ( |\alpha_0 \rangle+ |\beta_0 \rangle),
\end{equation}
 where $| \alpha \rangle \equiv \sum_{n=0}^{\infty} (\alpha^n / n !) \exp(-|\alpha|^2/2)  \hat{a}_0^{\dag n} |0 \rangle$ is a coherent state with complex amplitude $\alpha$ ($\alpha=\alpha_0, \beta_0$) and $\mathcal{N}$ is a normalization constant. In the following, we will typically assume $\beta_0=-\alpha_0$, so that $\mathcal{N}= 2+2 \exp(-2|\alpha_0|^2) \simeq 2$ for  a mean photon number $\langle n \rangle =|\alpha_0|^2$ larger than $\simeq 2$. Linear propagation in the lattice realizes a unitary map on the photon creation operators: the state vector of the system at a propagation distance $z$ is obtained from the expression of $| \mathcal{Q}(z=0) \rangle$ after the formal replacement $\hat{a}^{\dag}_0 \rightarrow \sum_{j=0}^{\infty} S_{j,0}(z) \hat{a}^{\dag}_j$, where $S_{j,0}(z)$ is the amplitude probability that one photon, initially injected into waveguide W ($j=0$), is found at waveguide $j$ after a propagation $z$ \cite{BO4,BO3,QW3}.  For a pure state, the density matrix of the full system $\{ÊW+B\}$ is given by $\hat{\rho}(z)=| \mathcal{Q} (z) \rangle \langle \mathcal{Q}(z) |$, and the reduced density matrix $\hat{\rho}_{W}(z)$ for the photon state in waveguide W can be readily obtained tracing over the degrees of freedom of the other waveguides. One obtains
\begin{eqnarray}
\hat{\rho}_W(z) & = & \frac{1}{\mathcal{N}} \left\{ | \alpha(z) \rangle \langle \alpha(z) |+| \beta (z) \rangle \langle \beta (z) | \right.  \\
& + & \left.  G(z)  | \alpha(z) \rangle \langle \beta(z) | +G^*(z)  | \beta(z) \rangle \langle \alpha(z) | \right\} \;\;\;\; \nonumber 
\end{eqnarray}
where we have set $\alpha(z)=\alpha_0 S_{0,0}(z)$, $\beta(z)=\beta_0 S_{0,0}(z)$ and
\begin{equation}
G(z)=\exp \left[-\frac{1}{2} \left( 1- |S_{0,0}(z)|^2 \right) (|\alpha_0|^2+|\beta_0|^2-2 \alpha_0^* \beta_0) \right].
\end{equation} 
Equation (4) clearly shows that the off-diagonal elements of the density matrix
$\hat{\rho}_W$ in the coherent-state basis $ \{ | \alpha(z) \rangle, | \beta(z) \rangle \}$ are damped by the factor $G(z)$, which 
is an exponential function of the single-photon decay probability $1-|S_{0,0}(z)|^2$, multiplied by $2 \langle n \rangle$ for $\beta_0=-\alpha_0$. The amplitude probability $S_{0,0}(z)$ can be computed by standard coupled-mode equation analysis \cite{Z1,Z2}. For $\sigma_0=0$ and in the weak coupling regime $\kappa_0 \ll \kappa$, one has $|S_{0,0}(z)|^2 \simeq \exp(-\gamma z)$ with $\gamma=2 \kappa_0^2 / \kappa$ \cite{Z1,Z2}, and hence the result given by Eqs.(4) and (5) exactly reproduces the ordinary model of decoherence for a Schr\"{o}dinger cat state in the dissipative quantum harmonic oscillator as obtained from the Lindblad master equation in the Markovian limit \cite{Harochebook,Mil}: the coherence rapidly decays toward zero with a characteristic lifetime given by $\sim 1/ (2 \gamma \langle n \rangle )$, which is extremely small for a macroscopic cat state (i.e. $\langle n \rangle \gg 1$). Non-Markovian features of the decoherence process arise from deviation of the single-photon survival probability $|S_{0,0}(z)|^2$ from an exponential function. As examples, in Fig.1(b) we plot the behavior of the coherence factor $G(z)$ for a Schr\"{o}dinger cat state with $\alpha_0=-\beta_0=3$ and a few values of $\kappa_0/ \kappa$ and $\sigma_0/ \kappa$. The figure clearly shows that, in the strong coupling regime, the decoherence process strongly deviates from the irreversible Markovian decay. In particular,  damped or sustained periodic revivals of the coherence  [curves 2 and 3 in Fig.1(b)] and fractional decoherence [curve 4 in Fig.1(b)] can be observed. Periodic revivals of the coherence, shown by curve 3 in Fig.1(b), are due to the existence of two non-degenerate bound states at the lattice edge (surface states), whose beating leads to the characteristic oscillating and non-decaying behavior of the single-photon survival probability. In case of curve 4 of Fig.1(b), the decoherence is fractional, i.e. after a transient the coherence factor $G(z)$ settles down to a steady-state and non-vanishing value. This is due to the existence of a single bound state, rather than to two non-degenerate bound states. This kind of  limited decoherence in the strong coupling regime was previously studied in Ref.\cite{frac0} and is found  in other systems as well, such as in  the spin-boson model \cite{frac}. As compared to the model of Ref.\cite{frac0}, in our case  fluctuations induced
from the coupling to the reservoir are not included.\par 
The results provided by Eqs.(4) and (5) for the decoherence of a Schr\"{o}dinger cat state that excites a reference waveguide W of the lattice are very general and hold for an {\it arbitrary} lattice structure.  To prove this statement, we propagate a quantized field in an arbitrary lattice structure beyond the tight-binding model using the rather general formalism of Ref.\cite{BO4}. Let us indicate by $u(x)$ the classical field profile of the guided mode sustained by the reference waveguide W of the lattice, with the normalization $\int dx |u(x)|^2=1$, and let us indicate by $\hat{a}^{\dag}_0=\int dx u(x) \hat{\phi}^{\dag}(x)$ the creation operator of photons in the waveguide mode, where $\hat{\phi}^{\dag}(x)$ is the bosonic creation operator of the monochromatic field satisfying the commutation relations $[\hat{\phi}(x), \hat{\phi}^{\dag}(x')]=\delta(x-x')$ and 
$[\hat{\phi}(x), \hat{\phi}(x')]= [\hat{\phi}^{\dag}(x), \hat{\phi}^{\dag}(x')]=0$ \cite{BO4}. Like in the previous example, the waveguide W is assumed to be excited at the input plane by a Schr\"{o}dinger cat state, i.e. the state vector of the field at $z=0$ is given by Eq.(3).  At a successive propagation distance $z$, the state vector $|\mathcal{Q}(z) \rangle$  of the photon field is obtained from its expression at $z=0$ after the formal substitution \cite{BO4} $\hat{a}^{\dag}_0 \rightarrow \int dx u(x,z) \hat{\phi}^{\dag}(x)$, where $u(x,z)$ is the classical field profile that propagates the initial field distribution $u(x,z=0)=u(x)$ in the lattice up to the distance $z$. $u(x,z)$ satisfies the classical wave equation, describing light propagation in the lattice, and can be numerically computed by standard methods. The classical field $u(x,z)$ can be rather generally decomposed as the superposition of the field that remains trapped in the waveguide W and the field that belongs to the other modes of the lattice (i.e. that has decayed into the bath), i.e. $u(x,z)=S_{0,0}(z) u(x)+ g(x,z)$, where $g(x,0)=0$, $S_{0,0}(0)=1$ and $\int dx u^*(x)g(x,z) \simeq 0$. Note that, like in the previous example,  $S_{0,0}(z)=\int dx u^*(x) u(x,z)$ represents the amplitude probability that one photon, initially injected into waveguide W at $z=0$, remains trapped in W after a propagation distance $z$. The reduced density matrix   $\hat{\rho}_{W}(z)$ for the photon state in waveguide W at the propagation distance $z$ is obtained from the full density matrix $\hat{\rho}(z)=| \mathcal{Q} (z) \rangle \langle \mathcal{Q}(z) |$ tracing over the degrees of freedom of the field that does not belong to the waveguide mode. After some lengthy but straightforward calculations, one obtains for   $\hat{\rho}_{W}(z)$  the expression given by Eqs.(4) and (5), which are thus of very general validity. In an experiment, the coherence of the cat state can be revealed by different means, for example by looking at the interference fringes in the distribution of a properly chosen field quadrature, from the evolution of the Wigner function or reconstructed off-diagonal density matrix elements, or from the modulation in the photon number distribution \cite{Harochebook,QE1,QE2}. The coherence factor $G$, defined by Eq.(5), is readily connected to such experimentally accessible quantities. For example, the photon number distribution in waveguide W at a given propagation distance $z$ for an initial cat with $\beta_0=-\alpha_0$ is given by $P_n(z)=\left( \rho_W(z)\right)_{n,n}= |\alpha(z)|^{2n} \exp(-|\alpha(z)|^{2n}) [1+(-1)^n G(z)]/(n !)$, which develops only along even number states for a coherent cat ($G=1$). The modulation in the photon number distribution is lacking in a statistical mixture ($G=0$), which contains all photon numbers and $P_n(z)=P_{n}^{(mix)}(z)$ is equal to the Poissonian distribution of the single coherent state $| \alpha \rangle$, which can be readily retrieved from a measure of $|\alpha(z)|^2$ with classical light. The coherence factor $G$ can be then estimated as $G(z)=P_n(z)/P_n^{(mix)}(z)-1$ for a fixed (even) value of $n$. Similarly, it can be shown that $G(z)$ is the visibility of fringes of the Wigner function (see, for instance, Eq.(9) of Ref.\cite{frac0}), which can be measured by quantum tomography methods. As an example, we discuss the decoherence of a Schr\"{o}dinger cat that undergoes BOs in a uniform waveguide lattice with a superimposed transverse index gradient \cite{BO1,BO2,BO3,BO4}. To account for the full band structure of the lattice and ZT, $S_{0,0}(z)$ is computed by numerical integration of the wave equation for the classical field $u(x,z)$, which in the scalar and paraxial approximations reads
\begin{figure}[htb]
\centerline{\includegraphics[width=8cm]{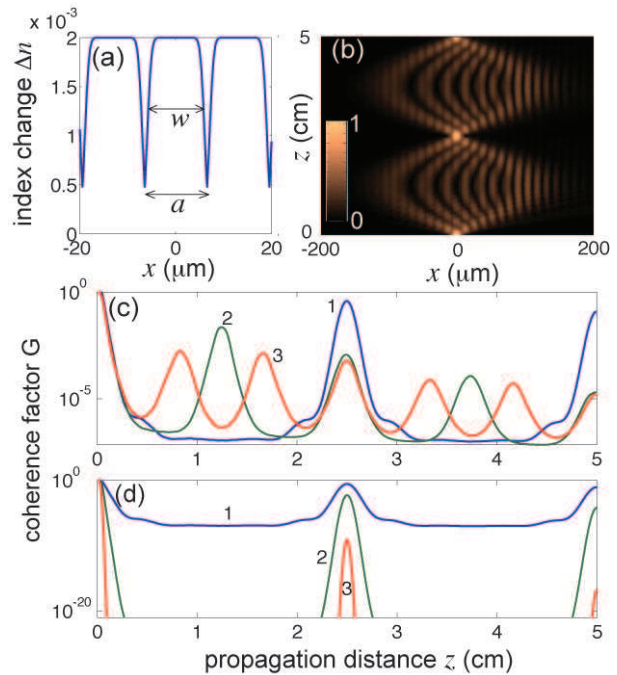}} \caption{
(Color online)  (a) Refractive index profile $\Delta n(x)$ of the array (lattice period $a=13 \; \mu$m, waveguide width $w=11.5 \; \mu$m). (b) Pseudocolor map showing the evolution of the classical field amplitude $|u(x,z)|$ (in arbitrary units) for an index gradient $F=3.61 \; {\rm cm}^{-1}$. The other parameter values are given in the text. (c) Evolution of the coherence  $G(z)$ (in logarithmic scale) for a 
a cat state with $\alpha_0=-\beta_0=3$ and for: $F=3.61 \; {\rm cm}^{-1}$ (curve 1), $F=7.22 \; {\rm cm}^{-1}$ (curve 2), and $F=10.83 \; {\rm cm}^{-1}$ (curve 3). (d) Evolution of $G(z)$ (in logarithmic scale) for $F=3.61 \; {\rm cm}^{-1}$ and for a cat state with increasing mean photon number: $\langle n \rangle =9$ (curve 1), $\langle n \rangle =36$ (curve 2), and $\langle n \rangle =144$ (curve 3).}
\end{figure}
$\partial_z u=-1/(2 k n_s) \partial^2_x u- k [ \Delta n(x)+Fx] u$,
where $k= 2 \pi / \lambda$ is the photon wave number, $n_s$ is the refractive index of the  dielectric substrate, $\Delta n(x)$ is the periodic modulation of refractive index with period $a$, and $F$ is the superimposed transverse index gradient, which is obtained for instance by circularly bending the waveguide axis \cite{R2,R3}.  As an example, in Fig.2  we show the results for the decoherence process of a Schr\"{o}dinger cat undergoing BOs and ZT. The simulations have been performed in a 5-cm-long array of waveguides with an index profile $\Delta n(x)$ shown in Fig.2(a),  a photon wavelength  $\lambda=1440$ nm, and a substrate refractive index $n_s=2.1381$; parameter values used in the simulations typically apply to Lithium-Niobate waveguide arrays \cite{DL}. Waveguide W is excited at the input plane in its fundamental mode. Figure 2(b) shows, as an example, the evolution of the classical field amplitude $|u(x,z)|$ along the array for an index gradient $F=3.61 \; {\rm cm}^{-1}$. The numerically-computed evolution of the coherence factor $G(z)$ is depicted in Fig.2(c) for $\alpha_0=-\beta_0=3$ and for a few increasing values of $F$. The behavior of $G(z)$ clearly shows periodic revivals, which are related to the onset of BOs with the photon probability distribution that initially spreads into adjacent waveguides due discrete diffraction \cite{R1} and  periodically returns into waveguide W with a spatial periodicity $z_B=\lambda/(Fa)$ \cite{R2}, see Fig.2(b). The revivals are imperfect owing to ZT, leading to a damping of the BOs with a damping factor that increases as the index gradient $F$ is increased. The damping of revivals in the coherence factor $G(z)$ are thus the signature of ZT that occurs in each BO cycle.  However, as compared to the damping of the photon density $|S_{0,0}|^2$, the damping of the coherence factor is strongly enhanced by the mean photon number $\langle n \rangle= |\alpha_0|^2$ of the cat state. Hence, even for a weak ZT, in a macroscopic cat state the decoherence of BOs is extremely fast and the density matrix rapidly reduces to an incoherent mixture of states $| \alpha(z) \rangle $ and $| \beta(z) \rangle $. This is clearly shown in Fig.2(d), where the evolution of the coherence factor $G(z)$ is plotted for a small index gradient $F$ and for increasing values of $\langle n \rangle$. \par
In conclusion, propagation of non-classical states of light (optical Schr\"{o}dinger cats) in waveguide lattices can provide an accessible and controllable laboratory tool to simulate decoherence processes with highly non-Marvokian features. Our suggested photonic system could provide a test bed to simulate with photons different dynamical control schemes of decoherence, based e.g. on ultrastrong coupling \cite{frac0,frac} or periodic modulation of the coupling to the continuum \cite{Kur}.

\newpage

\footnotesize {\bf References with full titles}\\
\\
\noindent
1. D. Christodoulides, F. Lederer, and Y. Silberberg, "Discretizing light behaviour in linear and nonlinear waveguide lattices",
Nature {\bf 424}, 817 (2003).\\
2. S. Longhi, "Quantum-optical analogies using photonic
structures", Laser and Photon. Rev. {\bf 3}, 243-261 (2009).\\
3. I.L. Garanovich, S. Longhi,  A.A. Sukhorukov, and Y.S. Kivshar, "Light propagation and localization in modulated photonic lattices and waveguides", Phys. Rep. {\bf 518}, 1 (2012).\\
4. U. Peschel, T. Pertsch, and F. Lederer, "Optical Bloch oscillations in waveguide arrays", Opt. Lett. {\bf 23}, 1701(1998).\\ 
5. H. Trompeter, T. Pertsch, F. Lederer, D. Michaelis, U. Streppel, A. Br\"{a}uer, and U. Peschel, "Visual observation of Zener tunneling", Phys. Rev. Lett. {\bf 96}, 023901 (2006).\\
6. S. Longhi, "Optical Bloch Oscillations and Zener Tunneling with Nonclassical Light", Phys. Rev. Lett. {\bf 101}, 193902 (2008).\\
7.Y. Bromberg, Y. Lahini, and Y. Silberberg, "Bloch oscillations of Path-Entangled Photons",  Phys. Rev. Lett. {\bf 105}, 263604 (2010).\\
8. S. Longhi, M. Marangoni, M. Lobino, R. Ramponi, P. Laporta, E. Cianci, and V. Foglietti, "Observation of dynamic localization in periodically-curved waveguide arrays", Phys. Rev. Lett. {\bf 96}, 243901 (2006).\\
9. Y. Lahini, A. Avidan, F. Pozzi, M. Sorel, R. Morandotti,
D. N. Christodoulides, and Y. Silberberg, "Anderson Localization and Nonlinearity in One-Dimensional Disordered Photonic Lattices", Phys. Rev. Lett.
{\bf 100}, 013906 (2008).\\
10. Y. Lahini, Y. Bromberg, D. N. Christodoulides, and Y. Silberberg, "Quantum Correlations in Two-Particle Anderson Localization",
Phys. Rev. Lett. {\bf 105}, 163905 (2010).\\
11. A. Crespi,	R. Osellame, R. Ramponi, V. Giovannetti,	 R. Fazio,	L. Sansoni, F. De Nicola,	F. Sciarrino, and
P. Mataloni, "Anderson localization of entangled photons in an integrated quantum walk", Nature Phot. {\bf 7}, 322 (2013).\\
12. S. Longhi, "Nonexponential Decay Via Tunneling in Tight-Binding Lattices and the Optical Zeno Effect", Phys. Rev. Lett. {\bf 97}, 110402 (2006).\\
13. F. Dreisow, A. Szameit, M. Heinrich, T. Pertsch, S. Nolte, A. T\"{u}nnermann, and S. Longhi, "Decay Control via Discrete-to-Continuum Coupling Modulation in an OpticalWaveguide System", Phys. Rev. Lett. {\bf 101}, 143602 (2008).\\
14. H.B. Perets, Y. Lahini, F. Pozzi, M. Sorel, R. Morandotti, and Y. Silberberg,
"Realization of quantum walks with negligible decoherence in waveguide lattices",
Phys. Rev. Lett. {\bf 100}, 170506 (2008).\\
15. A. Peruzzo, M. Lobino, J.C.F. Matthews, N. Matsuda, A. Politi,
K. Poulios, X.-Q. Zhou, Y. Lahini, N. Ismail, K. W\"{o}rhoff,
Y. Bromberg, Y. Silberberg, M. G. Thompson, and J.L. O'Brien, "Quantum Walks of Correlated Photons", Science {\bf 329}, 1500 (2010).\\
16. A. Rai, G. S. Agarwal, and J. H. H. Perk,
"Transport and Quantum Walk of Nonclassical Light in Coupled Waveguides", 
 Phys. Rev. A {\bf 78}, 042304 (2008).\\
17. L. Sansoni, F. Sciarrino, G. Vallone, P. Mataloni, A. Crespi, R. Ramponi, and R. Osellame, "Two-Particle Bosonic-Fermionic Quantum Walk via Integrated Photonics", Phys. Rev. Lett. {\bf 108}, 010502 (2012).\\ 
18. J.C.F. Matthews, K. Poulios, J.D.A. Meinecke, A. Politi, A. Peruzzo, N. Ismail, K. W\"{o}rhoff, M.G. Thompson, and J.L. O'Brien,
"Simulating quantum statistics with entangled photons: a continuous transition from bosons to fermions",  
Sci. Reports {\bf 3}, 1539 (2013).\\
19. W. Zurek, "Decoherence, einselection, and the quantum origins of the classical", 
Rev. Mod. Phys. {\bf 75}, 715 (2003).\\
20. S. Haroche and J.-M. Raimond, {\it Exploring the Quan-
tum: atoms, cavities and photons} (Oxford University
Press, Oxford, 2006).\\
21. S. Deleglise, I. Dotsenko, C. Sayrin, J. Bernu, M. Brune, J.-M. Raimond, and
S. Haroche, "Reconstruction of non-classical cavity field states
with snapshots of their decoherence", Nature {\bf 455}, 510 (2008).\\
22. H. Wang, M. Hofheinz, M. Ansmann, R. C. Bialczak, E. Lucero, M. Neeley, A. D. O'Connell, D. Sank, M. Weides,
J. Wenner, A.N. Cleland, and J.M. Martinis, "Decoherence Dynamics of Complex Photon States in a Superconducting Circuit", Phys. Rev. Lett. {\bf 103}, 200404 (2009).\\
23. K. H. Madsen, S. Ates, T. Lund-Hansen, A. L\"{o}ffler, S.
Reitzenstein, A. Forchel, and P. Lodahl, "Observation of Non-Markovian Dynamics of a Single Quantum Dot in a Micropillar Cavity", Phys. Rev. Lett.
{\bf 106}, 233601 (2011).\\
24. P. Huang, X. Kong, N. Zhao, F. Shi, P. Wang, X. Rong, R.-B. Liu, and J. Du, "Observation of an anomalous decoherence effect
in a quantum bath at room temperature", Nature Comm. {\bf 2}, 570 (2011).\\
25. W.-M. Zhang,P.-Y. Lo, H.-N. Xiong, M. W.-Y. Tu, and F. Nori, "General Non-Markovian Dynamics of Open Quantum Systems", Phys. Rev. Lett.Ê{\bf 109}, 170402 (2012).\\
26. A. Chiuri, C. Greganti, L. Mazzola, M. Paternostro, and P. Mataloni, "Linear Optics Simulation of Quantum Non-Markovian Dynamics", Sci. Rep. {\bf 2}, 968 (2012).\\
27. C.U. Lei and W.-M. Zhang, "Decoherence suppression of open quantum systems through a strong coupling to non-Markovian reservoirs", Phys. Rev. A {\bf 84}, 052116 (2011).\\
28. H.-B. Liu, J.-H. An, C. Chen, Q.-J. Tong, H.-G. Luo, and C.H. Oh, "Anomalous decoherence in a dissipative two-level system",  Phys. Rev. A {\bf 87},  052139 (2013).\\
29. J.T. Barreiro, M. M\"{u}ller, P. Schindler, D. Nigg,	 T. Monz, M. Chwalla, M. Hennrich, C.F. Roos, P. Zoller, and R. Blatt, "An open-system quantum simulator with trapped ions", Nature {\bf 470}, 486 (2011).\\
30. A. Ourjoumtsev, H. Jeong, R. Tualle-Brouri, and P. Grangier, "Generation of optical Schr\"{o}dinger cats from
photon number states", Nature {\bf 448}, 784 (2007).\\
31. D.F. Walls and G.J. Milburn, "Effect of dissipation on quantum coherence", Phys. Rev. A {\bf 31}, 2403 (1985).\\
32. A.G. Kofman and G. Kurizki, "Universal Dynamical Control of Quantum Mechanical Decay: Modulation of the Coupling to the Continuum", Phys. Rev. Lett. {\bf 87}, 270405
(2001).

\end{document}